\documentclass[twocolumn,showpacs,preprintnumbers,amssymb,aps,pra]{revtex4}
\usepackage{graphicx}
\usepackage{dcolumn}
\usepackage{bm}
\usepackage{latexsym,epsfig}

\usepackage{array}
\usepackage{amstext}

\begin{document}

\title{Conforming the measured lifetimes of the $5d \ ^2D_{3/2,5/2}$ states in Cs with theory}
\vspace*{0.5cm}

\author{B. K. Sahoo \footnote{Email: bijaya@prl.res.in}}
\affiliation{Theoretical Physics Division, Physical Research Laboratory, Ahmedabad-380009, India}

\date{Recieved date; Accepted date}
\vskip1.0cm

\begin{abstract}
\noindent
We find very good agreement between our theoretically evaluated lifetimes of the $5d \ ^2D_{3/2}$ and $5d \ ^2D_{5/2}$ states 
of Cs with the experimental values reported in [Phys. Rev. A {\bf 57}, 4204 (1998)], which were earlier evinced to be disagreeing 
with an earlier rigorous theoretical study [Phys. Rev. A {\bf 69}, 040501(R) (2004)] and with another precise measurement 
[Opt. Lett. {\bf 21}, 74 (1996)]. In this work, we have carried out calculations of the radiative transition matrix elements using 
many variants of relativistic many-body methods, mainly in the coupled-cluster theory framework, and analyze propagation of the 
electron correlation effects to elucidate their roles for accurate evaluations of the matrix elements. We also demonstrate 
contributions explicitly from the Dirac-Coulomb interactions, frequency independent Breit interaction and lower order quantum 
electrodynamics (QED) effects. Uncertainties to these matrix elements due to different possible sources of errors are estimated. 
By combining our calculated radiative matrix elements with the experimental values of the transition wavelengths, we obtain the 
transition probabilities due to both the allowed and lower order forbidden channels. Adding these quantities together, the 
lifetimes of the above two states are determined precisely and plausible reasons for the reported inconsistencies between the 
earlier theoretical calculations and the experimental results have been pointed out.
\end{abstract}

\maketitle

\section{Introduction}

Alkali atoms in general and Cs atom in particular are very interesting for many important experimental studies owing to their 
simple atomic energy level spacings and being suitable for optical magnetometry that are adequate to set up instrumentation 
for carrying out sophisticated measurements \cite{grimm,budker}. These atoms couple extremely weakly to the environment 
allowing potentially very long coherence times. Few such categorical experiments are quantum computing \cite{briegel,deutsch}, 
observing parity nonconservation (PNC) effects to probe new particle physics \cite{bouchiat,wood} and exotic property like 
nuclear anapole moment (NAM) \cite{haxton,ginges}, measuring electric dipole moment (EDM) due to parity and time reversal 
symmetry violations \cite{murthy,nataraj} etc.. In most of these studies, roles of accuracies in the theoretical calculations 
of various atomic properties are of also crucial \cite{ginges,nataraj,porsev,sahoo1,safronova,roberts1,roberts2}. In fact, one 
of the reasons why experiments on the alkali atoms are so popular because many calculations of the ground and excited states 
properties are already performed that are as precise as the experimental results. This provides good test of validity of both the 
experimental and theoretical results and entrust confidence to infer many fundamental physics by combining these results 
\cite{haxton,ginges,nataraj,porsev,sahoo1}. However, it is still found conflicts between the theoretical and experimental 
results of these atoms in few cases; especially while studying properties of the excited states. One of such examples is the 
disagreements between the calculations and experimental results of the lifetimes of the  $5d \ ^2D_{3/2}$ and $5d \ ^2D_{5/2}$ 
states of Cs \cite{safronova}. Again, there are at least two precise measurements have been carried out to determine the lifetimes 
of the $5d \ ^2D_{3/2}$ and $5d \ ^2D_{5/2}$ states of Cs \cite{diberardino,hoeling} among which the lifetime of the 
$5d \ ^2D_{5/2}$ state has been reported as 1281(9) ns \cite{diberardino} and 1225(12) ns \cite{hoeling}; which are clearly out 
of the ranges of their reported error bars. The difficult part to perform precise calculations of the excited state 
properties lies in the strong roles played by the continuum in the evaluation of the atomic wave functions. Moreover 
determination of wave functions of the excited states with large orbital angular momentum ($l$), e.g. $D$ states, demand large 
configuration state functions (CSFs) for which a method like configuration interaction (CI) method approximated only at the 
singles and doubles excitations (CISD method) may not be suitable enough to estimate their properties more accurately 
\cite{itano,sahoo2}. On the otherhand, relativistic coupled-cluster (RCC) method even at the same levels of singles and 
doubles approximation (CCSD method) seems to be capable of estimating many atomic properties within the reasonable accuracies
\cite{sahoo3,sahoo4,sahoo5}. One of the unique features of the (R)CC method is, it can  capture higher excited CSFs even 
approximating the method with the singles and doubles excitations owing to its ansatz of expressing the wave functions in the 
exponential form \cite{sahoo5,lindgren, szabo,bartlett}. For example, the CCSD method still accounts contributions from 
most of the triples, quadruples etc. CSFs through the non-linear terms \cite{szabo,bartlett}. Attaining precise results for 
the excited states with higher $l$ values, inclusion of the contributions from these triples and quadruples CSFs are 
imperative. However, the CCSD method with only the linear terms (LCCSD method) in addition with other corrections is often being 
employed in the atomic wave function calculations of the alkali atoms due to requirement of large computational resources for 
accounting for the non-linear RCC terms (e.g. see Ref. \cite{safronova} and references therein). Thus, comparative theoretical 
and experimental studies of the excited state atomic properties of the alkali atoms using a method would serve as a good test 
of its capability.

 This work is intended to probe again the validity of the reported experimental values of the lifetimes of the $5d \ ^2D_{3/2}$
and $5d \ ^2D_{5/2}$ states of Cs which, as mentioned above, are inconsistently reported in different works \cite{safronova}. 
Though experimental values from Refs. \cite{diberardino,hoeling} are discussed in Ref. \cite{safronova}, but few more measured 
values are also available in the literature with large error bars \cite{marek,sasso,bouchiat2}. Similarly few more calculations
of lifetimes of the above states are also known in the literature \cite{theodosiou, fabry, heavens, stone, warner}, however many 
of them are estimated using the non-relativistic theory and semi-empirical approaches. To appreciate consideration of the extra 
physical effects in the present work and to realize the reason for observing inconsistencies among the theoretical and 
experimental results, we discuss briefly here what was already done earlier in Ref. \cite{safronova}. In Ref. \cite{safronova}, 
the SD method, which is equivalent to the LCCSD method and SD method with important partial triples corrections (SDpT method) were
employed to calculate the electric dipole (E1) matrix elements. These values are further tweaked by scaling the wave functions of 
the SD method (SD$_{\text{sc}}$ method) and of the SDpT method (SDpT$_{\text{sc}}$ method). Using the E1 matrix elements 
from the SDpT method and separately combining with the values extracted from the measured lifetimes of the $5d \ ^2D_{3/2}$
and $5d \ ^2D_{5/2}$ states of Refs. \cite{diberardino} and with few more E1 matrix elements from the SDpT method, scalar dipole 
polarizabilities of the $6p \ ^2P_{1/2}$ and $6p \ ^2P_{3/2}$ states of Cs were evaluated. Differential polarizabilities of these 
$6P$ states with respect to the ground state were determined by taking the precisely measured ground state dipole polarizability 
as 401.0(6) $ea_0^3$ \cite{amini}, for the Bohr radius $a_0$. These differential polarizabilities were then compared with the 
direct measured values \cite{hunter1,tanner,hunter2}. This comparison demonstrated that the values obtained using the E1 matrix 
elements entirely from the SDpT method match better with the experimental results than the values obtained in the combined 
approach. On this basis it was argued that the measured values of the lifetimes of the $5D$ states and the differential 
polarizabilities of the $6P$ states were inconsistent. For this reason it was assumed that the E1 matrix elements obtained using 
the SDpT method are more accurate. Using the E1 matrix elements from the SD method of Ref. \cite{safronova}, the lifetimes of the 
$5d \ ^2D_{3/2}$ and $5d \ ^2D_{5/2}$ states were obtained as 1114 ns and 1547 ns respectively. These values were changed to 966 
ns and 1350 ns for the $5d \ ^2D_{3/2}$ and $5d \ ^2D_{5/2}$ states, respectively, when the matrix elements were improved using 
the SD$_{\text{sc}}$ method. Similarly, the lifetimes of the $5d \ ^2D_{3/2}$ and $5d \ ^2D_{5/2}$ states were obtained as 1010 ns 
and 1409 ns respectively upon the use of the matrix elements from the SDpT method. Finally, they were obtained as 981 ns and 
1369 ns respectively using the matrix elements from the SDpT$_{\text{sc}}$ method.

In the above analysis, two things need to be carefully scrutinized further. First, it can be noticed that after scaling the 
wave functions the results are changed significantly. This is very difficult to justify from the first principle accuracies in the 
results by scaling the wave functions. Large differences between the E1 matrix elements obtained before and after scaling the wave
functions indicate that it is necessary to consider more physical effects in the above employed SD or SDpT method to improve 
accuracies in the results. Secondly, it can be found from the sum-over-states approach employed in Ref. \cite{safronova} to 
evaluate polarizabilities of the $6P$ states that the principal contributions come from the E1 matrix elements between the 
$6p \ ^2P_{1/2} \rightarrow 5d \ ^2D_{3/2}$, $6p \ ^2P_{3/2} \rightarrow 5d \ ^2D_{3/2}$ and  $6p \ ^2P_{3/2} \rightarrow 
5d \ ^2D_{5/2}$ transitions. However, there are also ample amount of contributions come from the E1 matrix elements involving 
other $D$ excited states. Again, sum-over-states approach has limitations that it cannot account contributions from the core 
orbitals, higher excited states and continuum accurately. In order to verify accuracies of the theoretically estimated lifetimes 
of the $5D$ states, we would like to carry out calculations of the E1 matrix elements of the $6p \ ^2P_{1/2} \rightarrow 5d \ 
^2D_{3/2}$, $6p \ ^2P_{3/2} \rightarrow 5d \ ^2D_{3/2}$ and  $6p \ ^2P_{3/2} \rightarrow 5d \ ^2D_{5/2}$ transitions considering 
more physical effects than the LCCSD method; especially through the non-linear terms of the CCSD method. However, we also employ 
other lower order many-body methods such as the Dirac-Hartree-Fock (DHF) method, lower order perturbation theory (MBPT method) and 
LCCSD method to demonstrate gradual changes in the results with the propagation of the correlation effects through the higher 
order terms. We also give contributions from the important triples excitations in a variety of procedures and from a 
semi-empirical approach by using the experimental energies in the calculations of the wave functions in the RCC method. Again, the 
Dirac-Coulomb (DC) Hamiltonian was considered in Ref. \cite{safronova}. We also estimate corrections due to the higher order 
relativistic effects by considering the Breit interaction and lower order quantum electrodynamics (QED) effects in the 
calculations.

Many general implications for studying the lifetimes of the $5d \ ^2D_{3/2}$ and $5d \ ^2D_{5/2}$ states in Cs are already 
discussed before. We, however, would also like to emphasis on two applications here for which the present work could be directly 
relevant. The most precise PNC measurement has been carried out in the $6s \ ^2S_{1/2} \rightarrow 7s \ ^2S_{1/2}$ transition 
of Cs, but the nuclear parameters inferred from the NAM deduced from this measurement are in disagreement with the values given 
by the well established nuclear models \cite{haxton,ginges}. This urges for further investigation of PNC effects in the atomic 
systems. In fact, theoretical study demonstrates the PNC amplitudes in the $6s \ ^2S_{1/2} \rightarrow 5d \ ^2D_{3/2,5/2}$ 
transitions are almost three times larger than the $6s \ ^2S_{1/2} \rightarrow 7s \ ^2S_{1/2}$ transition in Cs 
\cite{roberts1,roberts2}. Plausible principle of measuring the PNC-induced frequency shift in the $6s \ ^2S_{1/2} \rightarrow 
5d \ ^2D_{3/2}$ transition of Ba$^+$ has been described in \cite{fortson}. Following this, it has also been suggested that the 
same principle can be adopted to measure PNC effect in the $6s \ ^2S_{1/2} \rightarrow 5d \ ^2D_{3/2}$ transition of Cs 
\cite{cho}. In another work, it has been highlighted that measurement induced light-shifts in the the $S-D_{5/2}$ transitions of 
the atomic systems would provide unambiguous signature of existence of NAM \cite{geetha,sahoo6}. One of the requirements to 
enhance the signal-to-noise ratio in the PNC-induced light-shift measurement principle is to have longer lifetimes of the states 
involved in the transition \cite{fortson}. Thus, it is indispensable to ensure reliability in the observed lifetimes of the  
$5d \ ^2D_{3/2}$ and $5d \ ^2D_{5/2}$ states in case the $6s \ ^2S_{1/2} \rightarrow 5d \ ^2D_{3/2,5/2}$ transitions in Cs are 
undertaken for the PNC measurement. On the otherhand,it was also advocated that the $5d \ ^2D_{3/2}$ and $5d \ ^2D_{5/2}$ states of Cs are very much suitable for the resonance 
ionization spectroscopy (RIS) process owing to their longer lifetimes \cite{hurst}. Therefore, it is important that ambiguity 
in the correctness of the lifetime values of the  $5d \ ^2D_{3/2}$ and $5d \ ^2D_{5/2}$ states of Cs are settled down. 

\section{Theory}

 In the Cs atom, it is obvious to assume that the dominant emission transition probabilities for an electron to jump from the 
$5d \ ^2D_{3/2}$ and $5d \ ^2D_{5/2}$ states are due to the E1 channel to the low-lying $6p \ ^2P_{1/2}$ and $6p \ ^2P_{3/2}$ 
stats. Since the aim of the present work is to explain the cause of disagreement between the previous theoretical calculations 
with the experimental results, we intend to show how much the transition probabilities are really small due to the forbidden 
channels from the above two states. Thus, we also determine transition probabilities due to the next dominant magnetic dipole (M1)
and electric quadrupole (E2) channels from the $5d \ ^2D_{3/2}$ state to the ground state $6s \ ^2S_{1/2}$ and from the 
$5d \ ^2D_{5/2}$ state to the ground and $5d \ ^2D_{3/2}$ states. The general expressions for evaluating these transition 
probabilities between the $\vert \Psi_i \rangle \rightarrow \vert \Psi_f \rangle$ transition are given by 
\begin{eqnarray}
&& A^{E1}_{if} = \frac{2.0261\times 10^{-6}}{\lambda_{if}^3 g_i} S_{if}^{E1} \label{eqn5} \\
&& A^{M1}_{if} =  \frac{2.6971\times 10^{-11}}{\lambda_{if}^3 g_i} S_{if}^{M1} \label{eqn6}  \\
\text{and}  && \nonumber \\
&& A^{E2}_{if} = \frac{1.1195\times 10^{-22}}{\lambda_{if}^5 g_i} S_{if}^{E2} \label{eqn7}, \ \ \ \ \ \ \
\end{eqnarray}
where the quantity $S^O_{if} = \mid {\langle \Psi_i \vert \vert O \vert \vert \Psi_f \rangle} \mid^2$ is known as the line 
strength for the corresponding reduced matrix element $\mid {\langle \Psi_i \vert \vert O \vert \vert \Psi_f \rangle} \mid$ of a 
transition operator $O$. These quantities are given later in this paper in atomic unit (a.u.). In the above expressions, 
$g_i=2J_i+1$ is the degeneracy factor of the state $\vert \Psi_i \rangle$ with the angular momentum of the state $J_i$ and the 
transition wavelength ($\lambda_{if}$) is used in nm which when substituted the transition probabilities ($A^O_{if}$s) are 
obtained in $s^{-1}$. The lifetime ($\tau$) of the atomic state $\vert \Psi_i \rangle$ is determined (in $s$) by taking 
reciprocal of the total emission transition probabilities due to all possible channels. i.e. 
\begin{eqnarray}
\tau_i &=& \frac {1} {\sum_{O,f} A^{O}_{if}},
\label{eqn8}
\end{eqnarray}
where the summations over $O$ and $f$ correspond to all the decay channels and all the lower states respectively. 

The reduced matrix elements for the E1, M1 and E2 transition operators in terms of the single particle orbitals are given by
\begin{eqnarray}
\langle \kappa_f\, ||\,e1\,||\, \kappa_i \rangle &=& \frac{3}{k}  \langle \kappa_f\, ||\,C^{(1)}\,||\,\kappa_i \rangle 
  \int_0^{\infty} dr \big ( j_1(kr) \nonumber \\ && \times \left [P_f(r)P_i(r)+Q_f(r)Q_i(r) \right ] + j_2(kr) \nonumber \\
  && \times \big \{ \frac{\kappa_f-\kappa_i}{2} \left [ P_f(r)Q_i(r)+Q_f(r)P_i(r)  \right ] \nonumber \\
&& +\left [ P_f(r)Q_i(r)-Q_f(r)P_i(r)  \right ] \big \} \big ), 
\end{eqnarray}
  \begin{eqnarray}
\langle \kappa_f\, ||\,m1\,||\, \kappa_i \rangle &=& \frac{6}{\alpha k}  \frac {(\kappa_f+\kappa_i)}{2} 
\langle - \kappa_f\, ||\,C^{(1)}\,||\,\kappa_i \rangle \int_0^{\infty} dr \nonumber \\ 
&& \times j_1(kr) \ (P_f(r)Q_i(r)+Q_f(r)P_i(r)) \ \ \  \ \ \ \ 
\end{eqnarray}
and
\begin{eqnarray}
\langle \kappa_f\, ||\,e2\,||\, \kappa_i \rangle &=& \frac{15}{k^2}  \langle \kappa_f\, ||\,C^{(2)}\,||\,\kappa_i \rangle 
  \int_0^{\infty} dr \big ( j_2(kr) \nonumber \\ && \times \left [P_f(r)P_i(r)+Q_f(r)Q_i(r) \right ] + j_3(kr) \nonumber \\
&& \times \big \{ \frac{\kappa_f-\kappa_i}{3} \left [ P_f(r)Q_i(r)+Q_f(r)P_i(r)  \right ] \nonumber \\
&& +\left [ P_f(r)Q_i(r)-Q_f(r)P_i(r)  \right ] \big \} \big ), 
\end{eqnarray}
where $P(r)$ and $Q(r)$ denote for the large and small components of the radial parts of the single particle Dirac orbitals, 
respectively, $\kappa$s are their relativistic angular momentum quantum numbers, $\alpha$ is the fine structure constant,
$k=\alpha (\epsilon_f - \epsilon_i)$ with the orbital energies $\epsilon$s and $j_l(kr)$ is the spherical Bessel function. 
The reduced Racah coefficients with rank $k$ are given by
\begin{eqnarray}
\langle \kappa_f\, ||\, C^{(k)}\,||\, \kappa_i \rangle &=& (-1)^{j_f+1/2} \sqrt{(2j_f+1)(2j_i+1)} \ \ \ \ \ \ \ \ \nonumber \\
                  &&  \left ( \begin{matrix} {
                          j_f & k & j_i \cr
                          1/2 & 0 & -1/2 \cr }
                         \end{matrix}
                            \right ) \pi(l_{\kappa_f},k,l_{\kappa_i}), \ \ \ \ \
\end{eqnarray}
with
\begin{eqnarray}
\pi(l,k,l') &=&
\left\{\begin{array}{ll}
\displaystyle
1 & \mbox{for } l+k+l'= \mbox{even}
\\ [2ex]
\displaystyle
0 & \mbox{otherwise,}
\end{array}\right.
\label{eqn12}
\end{eqnarray}
for the orbital momentum $l_{\kappa}$ of the corresponding orbital having the relativistic quantum number $\kappa$.

\begin{table}[t]
\caption{List of different parameters used to define the basis functions using QTOs in the present calculations.}
 \begin{ruledtabular}
  \begin{tabular}{lccccc}
   &  $s$ & $p$ & $d$& $f$ & $g$  \\
  \hline
 & & & \\
 $N_l$  & 34 & 33 & 32 & 31 & 30 \\
 $\eta_0$ & $2.0 \times 10^{-8}$ &  $2.5 \times 10^{-8}$ & $2.5 \times 10^{-8}$ & $2.1 \times 10^{-1}$ & $2.1 \times 10^{-7}$ \\
 $\zeta$ & 4.67 & 4.78 & 4.93 & 7.08 & 8.25 \\
  \end{tabular}
 \end{ruledtabular}
 \label{tab1}
\end{table}
For the first time, we use a different type of analytical basis function having quadratic type of exponents to express the single 
particle wave functions (define as quadratic type orbitals (QTOs)) to calculate the above reduced matrix elements. Using these 
functions, the radial components of the orbitals are expressed as
\begin{eqnarray}
P(r) \rangle &=& \sum_{\nu =1}^{N_l} c_{\nu}^P {\cal N}_{\nu}^P  r^l e^{-\eta_{\nu} r^4} \nonumber \\
\text{and} \ \ Q(r) \rangle &=& \sum_{\nu =1}^{N_l} c_{\nu}^Q  {\cal N}_{\nu}^Q  r^l  \left ( \frac{d}{dr} + \frac{\kappa}{r} \right ) \left [ r^l e^{-\eta_{\nu} r^4 } \right ] , \ \ \ \
\label{anbas}
\end{eqnarray}
where $N_l$ represents for the total number of QTOs considered in the calculations, $\eta_{\nu}$ is an arbitrary coefficient 
suitably chosen to obtain wave functions accurately, $c_{\nu}^{P(Q)}$s are the linear combination coefficients, 
${\cal N}_{\nu}^{P(Q)}$ is the normalization constant of the $\nu$th basis function for the large (small) component of the wave 
function. It can be noticed above that the kinetic balance condition between the large and small components has been maintained. 
The normalization constants are given by
\begin{eqnarray}
  {\cal N}_{\nu}^P &=& 2 (2 \eta_{\nu})^{\frac{2l+1}{8}} \Gamma \left ( \frac{2l + 1}{4}\right )^{-1/2}  
\end{eqnarray}
and
\begin{eqnarray}
 {\cal N}_{\nu}^Q &=& \Huge [ \frac{(l+\kappa)^2}{4(2 \eta_{\nu})^{\frac{2l-1}{4}}} \Gamma \left ( \frac{2l - 1}{4}\right )  - \frac{2(l+\kappa)}{(2 \eta_{\nu})^{\frac{2l-1}{4}}}  
 \nonumber \\ && \times \Gamma \left ( \frac{2l + 3}{4} \right ) +  \frac{4}{(2 \eta_{\nu})^{\frac{2l-1}{4}}} \Gamma \left ( \frac{2l + 7}{4}\right )  \Huge ]^{-1} . \ \ \
\end{eqnarray}
For convenience, the $\eta_{\nu}$ parameters are constructed satisfying the even tempering condition between two parameters 
$\eta_0$ and $\zeta$ as 
\begin{eqnarray}
\eta_{\nu} &=& \eta_0 \zeta^{\nu-1}.
\label{evtm}
\end{eqnarray}
We give the list of $\eta_0$ and $\zeta$ parameters in Table \ref{tab1} that are used in the present calculations. 

\section{Many-body methods}

In our previous work \cite{sahoo3}, we have described general procedures of our MBPT(2) and RCC methods using which we have 
calculated the wave functions and transition matrix elements of the Fr atom in the approach of Bloch's formalism 
\cite{lindgren}. We also adopt these methods here along with few more variants of the RCC methods by approximating the levels 
of excitations and non-linear terms in the expression of the wave function. We apply these methods in order to investigate 
the reason for the discrepancies between the previous theoretical study with the experimental results \cite{safronova}. We 
discuss briefly about these methods below to illustrate distinctly the roles of higher order correlation effects to enhance 
accuracies in the calculations of the transition matrix elements. 

In the Bloch's prescription the atomic wave function of a state $\vert \Psi_v \rangle$ of Cs with a valence orbital $v$ is 
expressed as \cite{lindgren}
\begin{eqnarray}
 \vert \Psi_v \rangle = \Omega_v \vert \Phi_v \rangle,
\end{eqnarray}
where $\Omega_v$ and $\vert \Phi_v \rangle$ are referred to as the wave operator and the reference state respectively. For the 
computational simplicity we choose the working reference state as the DHF wave function $\vert \Phi_c \rangle$ for the 
closed-shell configuration $[5p^6]$, which is common to the ground and the excited states that are involved in the estimations 
of the lifetimes of the $5D$ states of Cs. Then, the actual reference state is constructed from it as $\vert \Phi_v \rangle= 
a_v^{\dagger} \vert \Phi_c \rangle$ for the respective state with the valence orbital $v$. First the calculations are performed 
using the DC Hamiltonian which in a.u. is given by
\begin{eqnarray}
H &=& \sum_i \left [ c\mbox{\boldmath$\alpha$}_i\cdot \textbf{p}_i+(\beta_i -1)c^2 + V_n(r_i) + \sum_{j>i} \frac{1}{r_{ij}} \right ], \ \ \ \ \ \
\end{eqnarray}
with $\mbox{\boldmath$\alpha$}$ and $\beta$ are the usual Dirac matrices and $V_n(r)$ represents for the nuclear potential. We 
evaluate the nuclear potential considering the Fermi-charge distribution defined by
\begin{equation}
\rho_{n}(r)=\frac{\rho_{0}}{1+e^{(r-b)/a}},
\end{equation}
for the normalization factor $\rho_0$, the half-charge radius $b$ and $a= 2.3/4(ln3)$ is related to the skin thickness. We have 
used $a= 2.3/4(ln3)$ and $b=5.6707$ fm, which is  determined using the relation 
\begin{eqnarray}
b&=& \sqrt{\frac {5}{3} r_{rms}^2 - \frac {7}{3} a^2 \pi^2}
\end{eqnarray}
with the root mean square (rms) charge radius of the nucleus determined using the formula
\begin{eqnarray}
 r_{rms} =0.836 A^{1/3} + 0.570
\end{eqnarray}
in $fm$ for the atomic mass $A$.

Contributions from the frequency independent Breit interaction are estimated by adding the corresponding interaction term given by
\begin{eqnarray}
V_B(r_{ij}))=-\frac{1}{2r_{ij}}\{\mbox{\boldmath$\alpha$}_i\cdot \mbox{\boldmath$\alpha$}_j+
(\mbox{\boldmath$\alpha$}_i\cdot\bf{\hat{r}_{ij}})(\mbox{\boldmath$\alpha$}_j\cdot\bf{\hat{r}_{ij}}) \} .
\end{eqnarray}

We have also estimated lower order quantum electrodynamic (QED) effects by considering the following potentials with $H$ in a 
similar formalism as described in Ref. \cite{flambaum} but for the above nuclear Fermi-charge distribution. The lower order vacuum 
polarization (VP) effects are considered at the approximations of Uehling ($V_{U}(r)$) and Wichmann-Kroll ($V_{WK}(r)$) potentials 
given by
\begin{eqnarray}
V_{U}(r)&=&  - \frac{2 \alpha^2 }{3 r} \int_0^{\infty} dx \ x \ \rho_n(x) 
\int_1^{\infty}dt \sqrt{t^2-1} \nonumber \\ && \times 
\left(\frac{1}{t^3}+\frac{1}{2t^5}\right)  \left [ e^{-2ct|r-x|} - e^{-2ct(r+x)} \right ] \ \ \
\end{eqnarray}
and
\begin{eqnarray}
V_{WK}(r)&=&-\frac{8 Z^2 \alpha^4 }{9 r} (0.092) \int_0^{\infty} dx \ x \ \rho_n(x)  \nonumber \\ && \times \big ( 0.22 
\big \{ \arctan[1.15(-0.87+2c|r-x|)] \nonumber \\ && - \arctan[1.15(-0.87+2c(r+x))] \big \} \nonumber \\ && + 0.22 
\big \{ \arctan[1.15(0.87+2c|r-x|)] \nonumber \\ && - \arctan[1.15(0.87+2c(r+x))] \big \} \nonumber \\ 
&& - 0.11 \big \{ \ln[0.38 -0.87c|r-x|+c^2 (r-x)^2 ] \nonumber \\ 
&& - \ln[0.38 -0.87c (r+x) + c^2 (r+x)^2 ] \big \} \nonumber \\
&& + 0.11 \big \{ \ln[0.38 +0.87 c |r-x| + c^2 (r-x)^2 ] \nonumber \\ 
&& -  \ln[0.38 +0.87c (r+x) + c^2 (r+x)^2 ] \big \} \big ), \ \ \ \ \
\end{eqnarray}
with the atomic number of the system $Z$. The contribution from the self-energy (SE) interaction are accounted by evaluating 
contributions together from the electric form-factor given by
\begin{eqnarray}
V_{SE}^{ef}(r)&=& - A(Z) (Z \alpha )^4 e^{-Zr} + \frac{B(Z,r) \alpha^2 }{ r} \int_0^{\infty} dx  x  \rho_n(x) 
\nonumber \\ && \times \int^{\infty}_1 dt \frac{1}{\sqrt{t^2-1}} \big \{ \left( \frac{1}{t}-\frac{1}{2t^3} \right )\nonumber \\
&&\times \left [ \ln(t^2-1)+4 \ln \left ( \frac{1}{Z \alpha } +\frac{1}{2} \right ) \right ]-\frac{3}{2}+\frac{1}{t^2} \big \} \nonumber \\
&& \times \left [ e^{-2ct|r-x|} - e^{-2ct(r+x)} \right ]
\end{eqnarray}
and from the magnetic form-factor given by
\begin{eqnarray}
V_{SE}^{mg} (r) &=& \frac{i \alpha }{4 \pi c} \mbox{\boldmath$\gamma$} \cdot \mbox{\boldmath$\nabla$}_r \int_0^{\infty} d^3 x \ \rho_n(x) \nonumber \\
&& \times \left [ \left ( \int^{\infty}_{1}dt \frac{e^{-2tc R }}{Rt^2 \sqrt{t^2-1}}\right ) - \frac{1}{R} \right],
\end{eqnarray}
where $A(Z)=0.074+0.35Z \alpha $, $B(Z,r)=[1.071-1.97((Z-80) \alpha )^2 -2.128 ((Z-80) \alpha )^3+0.169 
((Z-80) \alpha )^4 ]cr/(cr+0.07(Z \alpha )^2 )$ and $R= | \textbf{r} - \textbf{x}|$. 
 
 Following the form of the reference states in our approach, $\Omega_v$ can now be divided as
\begin{eqnarray}
 \Omega_v =  1+ \chi_c + \chi_v ,
\end{eqnarray}
where $\chi_c$ and $\chi_v$ are responsible for carrying out excitations from $\vert \Phi_c \rangle$ and $\vert \Phi_v \rangle$, 
respectively, due to the residual interaction $V_r=H-H_0$ for the DHF Hamiltonian $H_0$. In a perturbative series expansion, 
we can express as
\begin{eqnarray}
 \chi_c = \sum_k \chi_c^{(k)} \ \ \text{and} \ \ \chi_v=\sum_k \chi_v^{(k)},
\end{eqnarray}
where the superscript $k$ refer to the number of times $V_r$ is considered in the MBPT method (denoted by MBPT(k) method). The 
$k$th order amplitudes for the $\chi_c$ and $\chi_v$ operators are obtained by solving the equations \cite{lindgren}
\begin{eqnarray}
 [\chi_c^{(k)},H_0]P &=& Q V_r(1+ \chi_c^{(k-1)} )P 
\end{eqnarray}
and
\begin{eqnarray}
[\chi_v^{(k)},H_0]P &=& QV_r (1+ \chi_c^{(k-1)}+ \chi_v^{(k-1)}) P 
 - \sum_{m=1 }^{k-1}\chi_v^{(k-m)} \nonumber \\ && \times PV_r(1+\chi_c^{(m-1)}+\chi_v^{(m-1)})P 
 \label{mbsv}
\end{eqnarray}
with $\chi_c^{(0)}=0$ and $\chi_v^{(0)}=0$, where the projection operators $P=\vert \Phi_c \rangle \langle \Phi_c \vert $ and 
$Q= 1- P$ describe the model space and the orthogonal space of the DHF Hamiltonian $H_0$ respectively. The energy of the state 
$\vert \Psi_n \rangle$ is evaluated by using an effective Hamiltonian
\begin{eqnarray}
 H_v^{eff}= P a_v H\Omega_v a_v^{\dagger} P.
 \label{efhm}
\end{eqnarray}
Using normal order Hamiltonian $H_N= H - PHP$ in place of $H$ in the above expression, attachment energy of a state with the 
valence orbital $v$ is evaluated. 

In the (R)CC theory ansatz, wave functions of the considered states are expressed as  
\begin{eqnarray}
 \vert \Psi_v \rangle & \equiv & \Omega_v \vert \Phi_v \rangle = e^T \{ 1+ S_v \} \vert \Phi_v \rangle
 \label{eqcc}
\end{eqnarray}
with $\chi_c= e^T-1$ and $\chi_v=e^TS_v -1$, where $T$ and $S_v$ are the CC excitation operators that excite electrons from the 
core and core along with the valence orbitals to the virtual space respectively. In this work, we have considered only the 
single and double excitations, denoted by the subscripts $1$ and $2$ respectively, in the CCSD method as
\begin{eqnarray}
 T=T_1 +T_2 \ \ \ \text{and} \ \ \ S_v = S_{1v} + S_{2v}.
\end{eqnarray}
In the LCCSD method only the linear terms are retained as in the SD method of Ref. \cite{safronova}. The amplitudes of these 
operators are evaluated using the equations
\begin{eqnarray}
 \langle \Phi_c^* \vert \overline{H}_N  \vert \Phi_c \rangle &=& 0
\label{eqt}
 \end{eqnarray}
and 
\begin{eqnarray}
 \langle \Phi_v^* \vert \big ( \overline{H}_N - \Delta E_v \big ) S_v \vert \Phi_v \rangle &=&  - \langle \Phi_v^* \vert \overline{H}_N \vert \Phi_v \rangle ,
\label{eqsv}
 \end{eqnarray}
where $\vert \Phi_c^* \rangle$ and $\vert \Phi_v^* \rangle$ are the excited state configurations, here up to doubles, with 
respect to the DHF states $\vert \Phi_c \rangle$ and $\vert \Phi_v \rangle$ respectively and $\overline{H}_N= \big ( H_N e^T 
\big )_l$ with subscript $l$ represents for the linked terms only. Here $\Delta E_v = H_v^{eff} - H_c^{eff}$ is the attachment 
energy of the electron in the valence orbital $v$ with $H_c^{eff}= P H \big ( 1+ \chi_c \big ) P$. Following Eq. (\ref{efhm}),
expression for $\Delta E_v$ is given by
\begin{eqnarray}
 \Delta E_v  = \langle \Phi_v \vert \overline{H}_N \left \{ 1+S_v \right \} \vert \Phi_v \rangle .
 \label{eqeng}
\end{eqnarray}

We also include contributions from the important triply excited configurations by defining perturbative operators defined as
\begin{eqnarray}
 T_{3}^{pert}  &=& \frac{1}{6} \sum_{abc,pqr} \frac{\big ( H_N T_2 \big )_{abc}^{pqr}}{\epsilon_a + \epsilon_b + \epsilon_c - \epsilon_p -\epsilon_q - \epsilon_r} ,
\label{t3eq}
 \end{eqnarray}
 and
\begin{eqnarray}
 S_{3v}^{pert} &=& \frac{1}{4} \sum_{ab,pqr} \frac{\big ( H_N T_2 + H_N S_{2v} \big )_{abv}^{pqr}}{\epsilon_a + \epsilon_b + \epsilon_v - \epsilon_p -\epsilon_q - \epsilon_r} ,
\label{s3eq}
 \end{eqnarray}
where $\{a,b,c \}$ and $\{ p,q,r \}$ represent for the occupied and virtual orbitals respectively and $\epsilon$s are their 
corresponding orbital energies. Since the final results reported in Ref. \cite{safronova} are using the SDpT method and scaling 
the wave functions, we would like to find out roles of the triply excited configurations in the evaluation of the transition
matrix elements. However the exact procedure using which triple excitations are accounted in the SDpT method is not clear to
us, so we try to estimate these contributions in various possible ways. When the $S_{3v}^{pert}$ operator is considered as a part 
of the $S_v$ operator to estimate only the energies using Eq. (\ref{eqeng}) after obtaining the RCC amplitudes, it is referred 
to as (L)CCSD(T) method. However, when it is involved to estimate both the energies and amplitudes of the $S_v$ operators 
simultaneously in the iterative procedure through Eqs. (\ref{eqsv}) and (\ref{eqeng}), we call it as (L)CCSD[T] method. To explore
roles of the core correlations through the triple excitations, we consider $T_3^{pert}$ operator as a part of $T$ operator while
solving Eq. (\ref{eqt}). This is referred to as (L)CCSDpT$^c$ method and when along with this approach, $S_{3v}^{pert}$ operator 
is considered in Eqs. (\ref{eqsv}) and (\ref{eqeng}), we refer to this as (L)CCSDpT method. But, we consider both the $T_3^{pert}$ 
and $S_{3v}^{pert}$ operators in Eqs. (\ref{eqt}) and (\ref{eqsv}) only to ameliorate amplitudes of the $T_1$ and $S_{1v}$ 
operators for the computational easiness.  

\begin{table}[t]
\caption{Demonstration of trends of the calculated energies (in cm$^{-1}$) using various relativistic methods considered in the 
present work with the DC Hamiltonian. Relativistic corrections are given separately from the CCSD method. These results 
are compared with the experimental values \cite{moore}. Uncertainties in the experimental values are not mentioned as they are 
more precise than the quoted values up to the second decimal places. Bold fonts are to highlight accuracies in the results.}
\begin{ruledtabular}
\begin{tabular}{lccccc} 
 Method  & $6s \ ^2S_{1/2}$ & $6p \ ^2P_{1/2}$ & $6p \ ^2P_{3/2}$ & $5d \ ^2D_{3/2}$ & $5d \ ^2D_{5/2}$ \\
 \hline
  & & \\
 DHF & 27983.73 & 18752.17 & 18350.36 & 14096.82 & 14121.80 \\
 MBPT(2) & 32020.63 & 20362.23 & 19777.17 & 16681.89 & 16568.09 \\
 LCCSD & 32425.64 & 20566.54 & 19965.95 & 17882.53 & 17718.94 \\
 LCCSD(T) & 31812.61 & 20335.35 &  19762.15 & 17439.99 & 17325.21 \\
 LCCSDpT$^c$ & 32425.66 & 20566.54 & 19965.95 & 17882.53 & 17718.94 \\
 LCCSD[T] & 31834.43 & 20340.48 &  19766.15 & 17505.35 & 17381.71 \\
 LCCSDpT  & 31758.84 & 20310.25 &  19741.51 & 17374.41 & 17281.31 \\
 {\bf CCSD} & 31463.22 & 20159.54 & 19600.28 & 16537.86 & 16445.08 \\
 CCSD(T) & 31090.05 & 20011.04 &  19470.48 & 16259.53 & 16149.22  \\
 CCSDpT$^c$ & 31428.69  & 20149.29 & 19591.82 & 16504.24 & 16414.53 \\
 CCSD[T] & 31064.13 & 20013.07 & 19472.02 & 16272.98 & 16214.82 \\
 CCSDpT & 31001.99  & 19993.26 & 19455.22 & 16223.73 & 16170.23  \\
 & & \\
 \multicolumn{6}{c}{Relativistic corrections } \\
  & & \\
  Breit & $-0.40$ & $-7.50$ & $-1.32$ & 20.17 & 23.62 \\
  VP   & 3.63 & $-0.03$ & $-0.09$ & $-0.40$ & $-0.36$ \\
  SE   & $-17.92$ & $-1.09$ & 0.95 & 2.11 & 2.15 \\
  Breit$+$QED & $-14.86$ & $-8.62$ & $-0.47$ & 21.87 & 25.42 \\
  & & \\
 Experiment & 31406.47 & 20228.20 &  19674.26 & 16907.21 & 16809.62 \\
\end{tabular}
\end{ruledtabular}
\label{tab2}
\end{table}

\begin{table}[t]
\caption{Comparison of E1 reduced matrix elements (in a.u.) from various methods. Relativistic corrections are quoted separately 
and our recommended values are given as ``Reco''. Results from other recent calculations are also given.}  
\begin{ruledtabular}
\begin{tabular}{lccc} 
  Method  &  $5d_{3/2} \rightarrow 6p_{1/2}$  & $5d_{3/2} \rightarrow 6p_{3/2}$  & $5d_{5/2} \rightarrow 6p_{3/2}$ \\
 \hline
  & & \\
 DHF & 9.012 & 4.078 &  12.233 \\
 MBPT(2) & 7.535 & 3.404 & 10.273 \\
 LCCSD & 6.566 & 2.954 &  9.011 \\
 LCCSD$_{\text{t3}}$ & 6.569 & 2.952 & 9.015 \\
 LCCSDpT$^c$ & 6.566 &  2.954 & 9.011 \\
 LCCSD[T] & 6.472 & 2.909 &  8.899 \\
 LCCSDpT  & 6.687 & 3.009  &  9.137     \\
 LCCSD$_{\text{ex}}$ & 6.305 & 2.828 & 8.683 \\
 CCSD$^{(2)}$ & 7.292 &  3.291 &  9.931 \\
 CCSD$^{(4)}$ & 7.301 &  3.295 &  9.941  \\
 {\bf CCSD}$^{(\infty)}$ & 7.301 & 3.295 & 9.941 \\
 CCSD$_{\text{t3}}$ &  7.304 & 3.293 & 9.945 \\
 CCSDpT$^c$ & 7.326 & 3.307  &  10.018 \\
 CCSD[T] & 7.258 & 3.275 & 9.934 \\
 CCSDpT  & 7.357 & 3.320 & 10.056  \\
 CCSD$_{\text{ex}}$ & 7.348 & 3.318 & 10.050 \\
 & & \\
 \multicolumn{4}{c}{Relativistic corrections} \\
 Breit & $-0.009$ & $-0.005$ & 0.022 \\
 VP   & $\sim 0.0$ & $\sim 0.0$ & 0.039 \\
 SE   & $-0.001$ & $-0.001$ & 0.037 \\
 Breit$+$QED &  $-0.010$&  $-0.005$  & 0.020  \\
 & & \\
 \multicolumn{4}{c}{Estimated Uncertainties} \\ 
 Basis & 0.048 & 0.023 &  0.022\\
 Triples & 0.003 & 0.002 & 0.004 \\
 Scaling & 0.047 & 0.023 & 0.109 \\
 & & \\
 \multicolumn{4}{c}{Recommended values} \\
  Reco  & 7.291(67) & 3.288(33) & 9.961(111) \\
 \hline
 & & \\
  \multicolumn{4}{c}{From Ref. \cite{safronova}} \\ 
  & & \\
 DHF & 8.9784 &  4.0625 & 12.1865 \\
 MBPT(3) & 6.9231 & 3.1191 & 9.4545 \\
 SD & 6.5809 & 2.9575 & 9.0238 \\
 SD$_{\text{sc}}$ & 7.0634 & 3.1871 & 9.6588\\
 SDpT & 6.9103 & 3.1112 & 9.4541 \\
 SDpT$_{\text{sc}}$ & 7.0127 & 3.1614 & 9.5906 \\
  & & \\
 \multicolumn{4}{c}{From Ref. \cite{roberts1}} \\ 
 & & \\
 $\Sigma^{(2)}$ & 6.744 &  3.037  &  9.254 \\
 $\lambda \Sigma^{(2)}$ & 7.039 & 3.173 & 9.629 \\
 $\Sigma^{(\infty)}$ & 6.927 & 3.121 & 9.481 \\
 $\lambda \Sigma^{(\infty)}$ & 7.032 & 3.170 & 9.616 \\
\end{tabular}
\end{ruledtabular}
\label{tab3}
\end{table}

\begin{table*}[t]
\caption{Reduced matrix elements (in a.u.) due to the E2 and M1 transitions given from different methods. Relativistic corrections and Reco 
values along with the uncertainties are given at the end of the table. The most accurate calculations are highlighted by the bold fonts.}
\begin{ruledtabular}
\begin{tabular}{lccccc} 
  Method  &  \multicolumn{2}{c}{$5d_{3/2} \rightarrow 6s_{1/2}$}  & $5d_{5/2} \rightarrow 6s_{1/2}$  & \multicolumn{2}{c}{$5d_{5/2} \rightarrow 5d_{3/2}$} \\
\cline{2-3} \cline{5-6} \\
       &  M1  &  E2       &  E2  &  M1  &  E2  \\ 
  \hline
  & & \\
 DHF      &  $\sim 0.0$  &   43.844    &  53.707  &  1.549 &  44.287 \\
 MBPT(2)  &  $3.2 \times 10^{-5}$  &  34.287   &  42.217  &  1.549  & 29.127  \\
 LCCSD    &  $8.8 \times 10^{-5}$  &  31.149   &  38.642  &  1.547  & 23.774 \\
 LCCSD$_{\text{t3}}$ & $8.8 \times 10^{-5}$  & 31.165    &  38.623  &  1.547 & 23.782 \\
 LCCSDpT$^c$ &  $8.8 \times 10^{-5}$  & 31.149 &  38.642  &  1.547  & 23.774 \\ 
 LCCSD[T] & $8.9 \times 10^{-5}$  &  30.698   &  38.153  & 1.547   &  23.033 \\ 
 LCCSDpT  & $8.7 \times 10^{-5}$  &  31.979   &  39.541  & 1.548   &  24.497 \\
 LCCSD$_{\text{ex}}$ &  $9.1 \times 10^{-5}$  &  30.053  &  37.436  & 1.548  & 21.741\\
 CCSD$^{(2)}$ & $7.6 \times 10^{-5}$  & 35.301  &  43.437  & 1.547  & 28.878 \\
 CCSD$^{(4)}$ & $2.1 \times 10^{-4}$  & 35.331  &  43.468  & 1.551  & 28.897 \\
 {\bf CCSD}$^{(\infty)}$ & $2.2 \times 10^{-4}$  & 34.400  &  42.441  & 1.551  & 29.037  \\
 CCSD$_{\text{t3}}$ & $2.2 \times 10^{-4}$  & 34.416 &  42.422 & 1.551 &  29.045 \\
 CCSDpT$^c$ &  $2.1 \times 10^{-4}$  &  34.523 &  42.598  &  1.551 &   29.248  \\
 CCSD[T] &  $2.2 \times 10^{-4}$  &  34.172  &  42.203  &  1.551   &  28.620 \\
 CCSDpT &  $2.0 \times 10^{-4}$ & 34.897 & 42.960 &  1.551 &  29.323 \\
 CCSD$_{\text{ex}}$ & $2.2 \times 10^{-4}$   &  34.530 & 42.600  & 1.551  &  29.622  \\
  & & \\
 \multicolumn{6}{c}{Relativistic corrections} \\
 & & \\
 Breit & $\sim 0.0$ & $-0.031$ & $-0.047$ & $\sim 0.0$ & $-0.105$ \\
 VP    & $\sim 0.0$ &  $-0.004$ & $-0.004$ & $\sim 0.0$ & 0.003 \\
 SE    & $\sim 0.0$  &  0.019 & 0.024  & $\sim 0.0$  & $-0.012$ \\
 Breit$+$QED & $\sim 0.0$  &  $-0.015$ & $-0.028$ &  $\sim 0.0$ &  $-0.116$ \\
 & & \\
 \multicolumn{6}{c}{Estimated Uncertainties} \\ 
 & & \\
 Basis &   $\sim 0.0$ &  0.106  &  0.112  &  $\sim 0.0$  &  0.289 \\
 Triples & $2.0 \times 10^{-5}$ & 0.016 & 0.019 & $0.001$  &  0.008  \\
 Scaling &  $\sim 0.0$ & 0.130 &  0.159 &  $\sim 0.0$  & 0.585  \\
  & & \\
 \multicolumn{6}{c}{Recommended values} \\
  & & \\
  Reco  &  $2.2(2) \times 10^{-4}$ & 34.385(168) & 42.413(195) &  1.550(1) & 28.921(653)  \\
\end{tabular}
\end{ruledtabular}
\label{tab4}
\end{table*}
  After obtaining amplitudes of the MBPT and RCC operators using the equations described earlier, the transition matrix element 
of an operator $O$ between the states $\vert \Psi_i \rangle$ and $\vert \Psi_f \rangle$ is evaluated using the expression
\begin{eqnarray}
\frac{\langle \Psi_f \vert O \vert \Psi_i \rangle}{\sqrt{\langle \Psi_f \vert \Psi_f\rangle \langle \Psi_i \vert \Psi_i\rangle}} 
&=& \frac {\langle \Phi_f \vert \Omega_f^{\dagger} O \Omega_i \vert \Phi_i\rangle}
{\sqrt{ \langle \Phi_f \vert \Omega_f^{\dagger} \Omega_f \vert \Phi_f\rangle \langle \Phi_i \vert \Omega_i^{\dagger} \Omega_i \vert \Phi_i\rangle} }  . \ \ \ \ \ 
\label{preq}
\end{eqnarray}
This gives rise to a finite number of terms for the MBPT(2) and LCCSD-variant methods, but it involves two non-terminating
series in the numerator and denominator, which are $e^{T^{\dagger}} O e^T$ and $e^{T^{\dagger}} e^T$ respectively, in the 
CCSD-variant methods. As described in our previous works \cite{sahoo3,yashpal1,yashpal2}, we adopt iterative procedures to 
account contributions from these non-truncative series. To comprehend, we also give intermediate results keeping different $k$ 
number of $T$ and/or $T^{\dagger}$ operators in these series of the CCSD method for evaluating the matrix elements and refer to 
the approach as CCSD$^{(k)}$ method. Finally, our CCSD results correspond to the calculations using the CCSD$^{(\infty)}$ method.
We also estimate contributions due to the triply excitations by considering both the $T_3^{pert}$ and $S_{3v}^{pert}$ operators 
along with their complex conjugates in Eq. (\ref{preq}) of the (L)CCSD method and refer the approach as (L)CCSD$_{\text{t3}}$ 
method.

\section{Results and Discussion}

Before presenting the transition matrix elements from various methods, we would like to first validate the methods by carrying 
out calculations of the attachment energies of the considered states of Cs and comparing them against their corresponding 
experimental values. Although it is understood that accuracies in the radial parts of the wave functions could be different in 
the accurate evaluation of the energies and transition matrix elements, but it can be noticed from Eqs. (\ref{mbsv}) and 
(\ref{eqsv}) that the energy evaluating expressions are also coupled with the wave function determining equations. Hence, accurate 
evaluation of the energies using a method can be an indication of the validation of the method in addition to embodying more 
physical effects in the method. For this purpose, we give energies obtained from various methods, that are described before, in 
Table \ref{tab2} using the DC Hamiltonian and compare them with the experimental values \cite{moore}. We find the CCSD method 
gives rise fairly accurate results for all the states being considered. It is also noticed that the MBPT(2) values are more 
accurate than the LCCSD values, but the partial triples effects bring down the LCCSD results closer to the experimental values.
However when these partial effects contributions are added in the CCSD results, the results become far off from the experimental 
results. It, therefore, implies that the neglected triples effects, mainly that can contribute through the $T_2$ and $S_{2v}$ 
amplitude determining equations, may cancel out some of these over estimated triple excited contributions to give finally more 
precise results. We also observe that the triples effects through the valence orbital excitations are the dominant ones over the 
core-triple excitations. Nevertheless, it would be pertinent to consider full triple excitations in this situation than adapting 
through the partial effects. Therefore, we consider results from the CCSD method, that accounts all the non-linear terms within 
the considered level of excitations, as the recommended calculated values for the further use.

We have also explicitly estimated the contributions due to the Breit interaction (given as ``Breit''), the VP effect (given as 
``VP'') and the SE effect (given as ``SE'') using the CCSD method which are given towards the bottom of Table \ref{tab2}. In 
addition, we also determine these corrections considering all these relativistic corrections together with respect to the 
contributions from the DC Hamiltonian in the CCSD method (given as ``Breit$+$QED''). We find slight changes in the results between
the ``Breit$+$QED approach and when the corrections estimated independently are added-up. We also observe that among all these 
relativistic corrections, the SE effect is large in the ground state while the Breit interaction gives larger corrections in the 
other considered states.

\begin{table*}[t]
\caption{Transition wavelengths (in nm) and probabilities ($A_{if}^O$) due to different decay channels ($O$s) in $s^{-1}$ 
from the $5d \ ^2D_{3/2}$ and $5d \ ^2D_{5/2}$ states of Cs from various works. Uncertainties are quoted within the 
parentheses. We also compare our results with the available other theoretical and experimental values. Results only from the 
SDpT$_{\text{sc}}$ method are quoted from Ref. \cite{safronova}.}
\begin{ruledtabular}
\begin{tabular}{lccccccc} 
   Transition             & $\lambda_{if}$ (nm) & $O$ & \multicolumn{2}{c}{$A_{if}^O $ (in $s^{-1}$)} & \multicolumn{3}{c}{$\tau_{i}$ (in ns)}  \\  
 \cline{4-5} \cline{6-8} \\
   $J_i \rightarrow J_f$  &    \cite{moore} &     &   This work  &  Others &   This work & Others & Experiment \\        
  \hline 
 & & \\ 
 $5d \ ^2D_{3/2} \rightarrow 6p \ ^2P_{1/2}$  & 3011.15   &  E1 &  986229(18209)  & 804000 \cite{safronova} &   907(16)  & 981 \cite{safronova} &  909(15) \cite{diberardino} \\
 $5d \ ^2D_{3/2} \rightarrow 6p \ ^2P_{3/2}$  & 3613.96   &  E1 &  116015(2341)   & 94000 \cite{safronova}  &            & 909 \cite{theodosiou} &  890(90) \cite{marek} \\
 $5d \ ^2D_{3/2} \rightarrow 6s \ ^2S_{1/2}$  & 689.69    &  E2 &  21.21(20)      &         &            & 1061 \cite{fabry} &  1250(115) \cite{sasso} \\
 $5d \ ^2D_{3/2} \rightarrow 6s \ ^2S_{1/2}$  &           &  M1 & $\sim 0$        &         &            & 952  \cite{heavens} &  \\
                                              &           &     &                 &         &            & 970 \cite{stone}  &  \\
                                              &           &     &                 &         &            & 856 \cite{warner}  &  \\
 \hline
 & & \\
 $5d \ ^2D_{5/2} \rightarrow 6p \ ^2P_{3/2}$  & 3490.84   &  E1 & 787636(17652)&  646000 &  1270(28) & 1369 \cite{safronova} &  1281(9) \cite{diberardino} \\
 $5d \ ^2D_{5/2} \rightarrow 6s \ ^2S_{1/2}$  & 685.08    &  E2 &  22.24(21)   &         &           & 1283 \cite{theodosiou}  &  1225(12) \cite{hoeling} \\ 
 $5d \ ^2D_{5/2} \rightarrow 5d \ ^2d_{3/2}$  & 102469.52 &  E2 & $\sim 0$     &         &           & 1434 \cite{fabry}   &  890(90) \cite{marek} \\
 $5d \ ^2D_{5/2} \rightarrow 5d \ ^2d_{3/2}$  &           &  M1 & $\sim 0$     &         &           & 1370 \cite{heavens} &  1250(115) \cite{sasso} \\
                                              &           &     &              &         &           & 1342 \cite{stone}  &   1260(80) \cite{bouchiat2} \\
                                              &           &     &              &         &           & 1190 \cite{warner} &   \\
\end{tabular}
\end{ruledtabular}
\label{tab5}
\end{table*}

After analyzing energies from various methods with respect to the experimental values, we also give the E1 matrix elements in 
Table \ref{tab3} of the transitions that are required to estimate the lifetimes of the $5d \ ^2D_{3/2}$ and $5d \ ^2D_{5/2}$ 
states of Cs. We compare these results with the values reported recently by the other groups using different relativistic 
many-body methods \cite{safronova,roberts1}. In addition to the methods we have employed to evaluate the energies, we also give 
E1 matrix elements using the (L)CCSD$_{\text{t3}}$ and (L)CCSD$_{\text{ex}}$ methods in the above table. We find reasonable 
agreements between the results obtained using our DHF and LCCSD methods with the DHF and SD methods of Ref. \cite{safronova}. 
However, there are significant differences in the results when the higher order effects are accounted. Similarly, our results 
differ substantially from the calculations reported in Ref. \cite{roberts1} in which a combination of correlation potential (CP)
method ($k$th order is denoted by $\Sigma^{(k)}$) and time-dependent Hartree-Fock (TDHF) method with the Brueckner orbitals (BOs) 
are employed. Moreover, the E1 matrix elements quoted in Ref. \cite{safronova} are improved drastically using the SD$_{\text{sc}}$ 
and SDpT$_{\text{sc}}$ methods where the wave functions are scaled to account the omitted contributions. Large differences 
in the results obtained before and after scaling the wave functions demand for including the omitted contributions more accurately. In Ref. \cite{roberts1}
too, the final results are quoted using the $\lambda \Sigma^{(k)}$ approach with the scaling parameter $\lambda$.  Our CCSD method 
includes more physical effects through its formulation \cite{lindgren,szabo,bartlett} and this is also partly justified from 
the comparison of energies in Table \ref{tab2}. Therefore, we consider results from the CCSD method as more reliable since it includes 
all the non-linear terms within the considered levels of approximations and accounts pair-correlation and core-polarization 
effects to all orders \cite{sahoo5}. These non-linear terms take care of most of the contributions from the triple and quadrupole 
excitations; more importantly both the singly and doubly excited amplitudes see these effects equitably. To show the effectiveness 
of these non-linear terms, we also evaluate the E1 matrix elements considering the same linear form of the RCC terms in 
Eq. (\ref{preq}) through our CCSD$^{(2)}$ method that naturally appears in the LCCSD method. As seen in Table \ref{tab3}, the 
differences between the results from the LCCSD and CCSD$^{(2)}$ methods are quite large. This countenances our above 
assertion. Compared to the results from the MBPT(2) method, results obtained using the (LCC)SD and MBPT(3) methods from
our calculations and from Ref. \cite{safronova} are smaller but the CCSD values are closer. This means there are strong 
cancellations in the correlation effects among the higher order terms. This trend is similar to the calculations of energies 
as seen in Table \ref{tab2}. We also notice amount of contributions estimated through the partial triples effects by us and given 
in Ref. \cite{safronova} are very different. This may be owing to the fact that triples effects are incorporated differently  
in both the works. Differences between the LCCSD and LCCSDpT results in our calculations are larger than the difference between
the CCSD and CCSDpT results. It means partial triples effects change results in the LCCSD approximation more than the CCSD 
method. In contrast, we find the differences between the LCCSD and LCCSDpT$^c$ results are much smaller than the differences 
between the CCSD and CCSDpT$^c$ results implying core-correlations enhance through the non-linear terms of the RCC theory. We 
also observe relatively small changes in the results obtained using the CCSD$^{(2)}$, CCSD$^{(4)}$ and CCSD$^{(\infty)}$ 
approximations. Thus, the roles of the non-linear terms of the CCSD method are more effective in the determination of the 
wave functions than the property evaluation. We also observe both the Breit and QED corrections are of decent size for
determining precise value of the E1 matrix element of the $5d \ ^2D_{5/2} \rightarrow 6p \ ^2P_{3/2}$ transition. In fact it 
is interesting to note here that, unlike in the energy calculations, total sum of the relativistic corrections to the above
E1 matrix element obtained from the Breit interaction, VP effect and SE effect are quite different than when they are estimated 
considering all the interactions (Breit$+$QED approach) together in the CCSD method. In the other transitions, these corrections 
are found to be mere in magnitude.

In order to satisfactorily address issues related to the inconsistencies between the previously estimated theoretical results for 
the lifetimes of the $5d \ ^2D_{3/2}$ and $5d \ ^2D_{5/2}$ states of Cs with the experimental values, it is also essential 
to estimate uncertainties associated with the E1 matrix elements carefully. Obviously, it can be argued that our calculations have 
uncertainties from three major sources: (a) use of finite basis size, (b) approximations in the levels of excitations in the RCC 
theory and (c) {\it ab initio} approach for calculating the wave functions. Among these three, the first two sources of 
uncertainties are quite understandable. To fathom about the uncertainty due to the {\it ab initio} approach, one can follow from 
Eq. (\ref{eqsv}) that both the wave function and energy determining equations are coupled. Therefore, uncertainties associated in 
both the solutions either may be canceled out each other or will be added-up in the final evaluation. If the experimental energy is used in
Eq. (\ref{eqsv}) (which may be teated as a semi-empirical approach) then the uncertainty associated with the energy can be 
removed (assuming that the experimental energy is more precise). We estimate uncertainties due to the truncated basis size 
(given as ``Basis'') by carrying out calculations with the high lying orbitals using the MBPT(2) method that are neglected in the RCC 
calculations to circumvent the computational limitations. Uncertainty due to the neglected triples (given as ``Triples'') are 
accounted by taking differences between the results obtained using the CCSD and CCSD$_{\text{t3}}$ methods. For estimating 
uncertainties due to the {\it ab initio} calculations (given as ``Scaling''), we consider differences between the results from 
the CCSD and CCSD$_{\text{ex}}$ methods. It is worth mentioning here that most of the partial triples effects seen in our 
calculations are present inherently within the above estimated differences. Therefore, it takes into account almost all possible 
major uncertainties of our calculations using the CCSD method. By accounting all these uncertainties in quadrature, we give the 
recommended values (quoted as ``Reco'') of the E1 matrix elements towards the end of Table \ref{tab3}. The absolute values are 
given after adding the relativistic corrections to the CCSD results.

 It can be assumed that contributions from the forbidden transition probabilities to the estimations of the lifetimes of the 
$5d \ ^2D_{3/2}$ and $5d \ ^2D_{5/2}$ states of Cs are negligibly small. However, it is necessary to demonstrated in the 
scenario when there are inconsistencies among the theoretical and experimental results. For this purpose, we also estimate these  
quantities explicitly for the lower order M1 and E2 forbidden channels. We give these forbidden transition amplitudes from the 
$5d \ ^2D_{3/2}$ and $5d \ ^2D_{5/2}$ states in Table \ref{tab4} using the same methods that were employed to calculate 
the E1 matrix elements. We also give uncertainties to these quantities adopting the same procedure described in the previous 
paragraph. The trends of these matrix elements from different many-body methods are almost similar to the E1 results except 
for the M1 matrix element between the $5d \ ^2D_{3/2} \rightarrow 6s \ ^2S_{1/2}$ transition; which is anyway found to be 
negligibly small. The relativistic corrections are also found to be quite small. The recommended values are given at the end of 
the table following the same procedure as were given for the E1 matrix elements in Table \ref{tab3}.

Using the recommended transition matrix elements given in Tables \ref{tab3} and \ref{tab4} and the experimental wavelengths, 
quoted in Table \ref{tab5} from the database \cite{moore}, we determine the transition probabilities due to all the considered 
channels from the $5d \ ^2D_{3/2}$ and $5d \ ^2D_{5/2}$ states of Cs. These values are quoted in Table \ref{tab5} along with 
their uncertainties and compared against the values due to the E1 channel obtained using the SDpT$_{\text{sc}}$ procedure of 
Ref. \cite{safronova}. We find large differences between the results from both the works. From the total polarizabilities of 
these results, we find the lifetime of the $5d \ ^2D_{3/2}$ state is 907(16) ns against 981 ns reported in Ref. \cite{safronova}. 
This is in quite good agreement with the experimental value 909(15) ns reported in Ref. \cite{diberardino}. Similarly, we obtain 
lifetime of the $5d \ ^2D_{5/2}$ state as 1270(28) ns against 1369 ns of Ref. \cite{safronova}. Our result again agrees well 
with the experimental value 1281(9) ns reported in Ref. \cite{diberardino}. In Table \ref{tab5}, we also quote estimated 
lifetimes of these states from some of the previous theoretical and other experimental results. Most of these theoretical 
estimations were carried out using the non-relativistic theory \cite{theodosiou, fabry, heavens, stone, warner}. Nevertheless, 
theoretically estimated values in Ref. \cite{theodosiou} are very close to our values and the experimental results of 
\cite{diberardino}. Other theoretical results are far away from our calculations. Other experimental values for the lifetimes 
of the above $5D$ states also have large error bars \cite{marek,sasso,bouchiat2} except for the $5d \ ^2D_{5/2}$ state as 
1225(12) ns reported in Ref. \cite{hoeling}. This lies outside the range of the error bar of the value reported in Ref. 
\cite{diberardino}.  Since we have overestimated the uncertainties in our theoretical analysis to provide more reliable results,
we anticipate that error bars in our calculations would be smaller than what have been actually reported. From this point of view,
our results support experimental values of the lifetimes of the $5D$ states of Cs reported in Ref. \cite{diberardino}. Our 
calculations also demonstrate that branching ratios of an electron to jump from the $5d \ ^2D_{3/2}$ state to the lower
$6p \ ^2P_{1/2}$ is about 90\% while to the $6p \ ^2P_{3/2}$ state is about 10\%. On the otherhand, an electron can jump from the
$5d \ ^2D_{5/2}$ to the $6p \ ^2P_{3/2}$ with almost 100\% probability.

\section{Conclusion}

We have employed a variety of relativistic many-body methods mostly in the coupled-cluster theory framework to calculate 
the energies and transition matrix elements due to both the allowed and forbidden decay channels of the $5d \ ^2D_{3/2}$ and 
$5d \ ^2D_{5/2}$ states in Cs. Trends in the results from these methods are discussed and importance of considering the 
non-linear terms for accurate determination of the matrix elements are highlighted. Corrections due to both the Breit 
interaction and lower order QED effects in these quantities are demonstrated explicitly. Earlier reported inconsistencies 
between the theoretical and experimental values of the lifetimes of the above states seem to be resolved. Branching ratios 
due to various channels are also given. Though the forbidden transition probabilities are found to be extremely small, however
our calculated values can be quite useful if the proposed measurements of parity non-conservation effects in the $6s \ ^2S_{1/2} 
\rightarrow 5d \ ^2D_{3/2,5/2}$ transitions in Cs take place in future.

\section*{Acknowledgement}

Computations were carried out using the Vikram-100TF HPC cluster at the Physical Research Laboratory, Ahmedabad, India.

\end{document}